\documentclass[twocolumn]{revtex4}
\usepackage{graphicx}
\usepackage[]{epsfig}

\newcommand{\beq}{\begin{equation}}
\newcommand{\eeq}{\end{equation}}
\newcommand{\beqd}{\begin{displaymath}}
\newcommand{\eeqd}{\end{displaymath}}
\newcommand{\beqa}{\begin{eqnarray}}
\newcommand{\eeqa}{\end{eqnarray}}

\newcommand{\comment}[1]{}

\begin{document}
\title{On $k$-Core Percolation in Four Dimensions}


\author{Giorgio Parisi$^{1}$ and Tommaso Rizzo$^{2}$}

\affiliation{$^{1}$Dipartimento di Fisica, Universit\`a di Roma ``La Sapienza'', 
P.le Aldo Moro 2, 00185 Roma,  Italy
\\
$^{2}$ ``E. Fermi'' Center, Via Panisperna 89 A, Compendio Viminale, 00184, Roma, Italy}

\begin{abstract}
The $k$-core percolation on the Bethe lattice has been proposed as a simple model of the jamming
transition because of its  hybrid first-order/second-order nature.
We investigate numerically  $k$-core percolation on the four-dimensional regular lattice. For $k=4$
the presence of a discontinuous transition is clearly established but its nature is strictly first order. In particular, the $k$-core density displays no singular behavior before the
jump and its correlation length remains finite. For $k=3$ the transition is continuous.
\end{abstract} 

\maketitle

After its introduction \cite{Chalupa}, $k$-core percolation has been proposed to be relevant in a
variety of different contexts, see \cite{Adler1, Adler2}.
The problem, also referred to as "Bootstrap percolation", is defined as follows. 
The sites of a given lattice are populated with probability $p$. Then each site with less than $k$ neighbors is
removed, the procedure being iterated until each site has at least $k$ neighbors.

On regular lattices in dimension $d$ the model exhibits two different behaviors depending on the
value of $k$.
If $k>d$ the cluster must be extended in order to survive the culling process but it is completely
decimated for any $p<1$ in the large size limit \cite{Schon}.
The behavior  for large but finite system size has also been investigated \cite{Aizenman, Holroyd, Cerf,
VanEnter} but strong disagreement between the theoretical predictions and numerical simulations has been
found. The highly non trivial origin of this discrepancy has been clarified only recently
\cite{DeGregorio1}.
If $k \leq d$ it is easy to realize that there are small "self-sustained" structures ({\it e.g.}
$d$-dimensional hypercubes) that can survive culling irrespective of their environment. In this
case the $k$-core always exists and the problem is rather if it percolates or not and
what is the nature of the percolation transition \cite{Harris2}.

On the Bethe lattice the $k$-core percolation transition is known to be discontinuous
\cite{Koguth}. Starting from high values of $p$ the density of the $k$-core drops discontinuously
to zero at $p_c$. The transition however is not simply first order, the density near the transition
is given by $\rho(p) \simeq \rho(p_c)+b (p-p_c)^{1/2}$.
Furthermore it has been recently pointed out that the transition  is also accompanied by diverging
correlation lengths.
This behavior has motivated the proposal of $k$-core percolation as a model of the jamming
transition \cite{Schwarz}. There are indeed evidences that this transition has a mixed
character \cite{jamm}. 

This brought new interest into the question 
of  wether the hybrid  nature of the transition in the Bethe lattice survives in finite
dimensions.
This is certainly not the case for cubic lattices in $d \leq 3$.
Indeed for $k=2$  the transition is continuous and has the same critical point of ordinary
percolation \cite{Harris1}.
For $k=3$ and $d=3$ the transition is continuous \cite{Koguth,Branco,Kurtsiefer,Medeiros} with
exponents consistent with those of ordinary $d=3$ percolation \cite{Branco}. 
These results are valid on cubic lattices and they not exclude the possibility of a mixed
transition for $d\leq 3$ provided the structure of the lattice or the constraints are different.
Indeed recently a $2$-dimensional model with  a mixed transition was exhibited \cite{Cris} and
numerical evidences of a mixed transition in another $2$-dimensional model were reported in
\cite{Schwarz}.
As for regular lattices, an expansion in powers of $1/d$ has proven that turning on dimension perturbatively
does not destroy the mixed nature of the transition \cite{Harris2}, thus suggesting that the hybrid
transition may exist for some $(d>3,2<k<d+1)$.
In this work we investigate numerically $k$-core percolation on the four-dimensional hypercubic
lattice.
For $k=3$ we have found a continuous transition and we did not further investigate the critical
behavior.
In the case $(d=4,k=4)$ we find negative results concerning the hybrid transition: while the
presence of a discontinuous transition is clearly established, it is strictly first order.
More precisely for $k=4$, at a critical value $p_c=.6885(5)$ the system has a phase transition from
a high $p$ phase where there is a giant cluster with a finite density to a low-$p$ phase where there is not.
The density of the giant cluster is given by the $k$-core density minus the density of the small clusters, that is approximately $\rho_{small} \approx 0.04$ near the transition, therefore the critical properties of the giant cluster can be safely extracted from the total density in the percolating phase. The $k$-core density exhibits a discontinuous transition, jumping from a $\rho_c^{+}=0.567(4)$ to
$\rho_c^{-}=0.044(2)$. 
However the density displays no singular behavior at the transition, and the correlation length
extracted from $k$-core correlation function $G(i,j)=\langle \nu_i \nu_j \rangle -\langle \nu_i
\rangle \langle \nu_j \rangle$ (where $\nu_1$ is $1$ on the $k$-core and $0$ otherwise
\cite{Harris2}) remains finite, $\xi_c<10$\, ;

\begin{figure}[htb]
\begin{center}
\epsfig{file=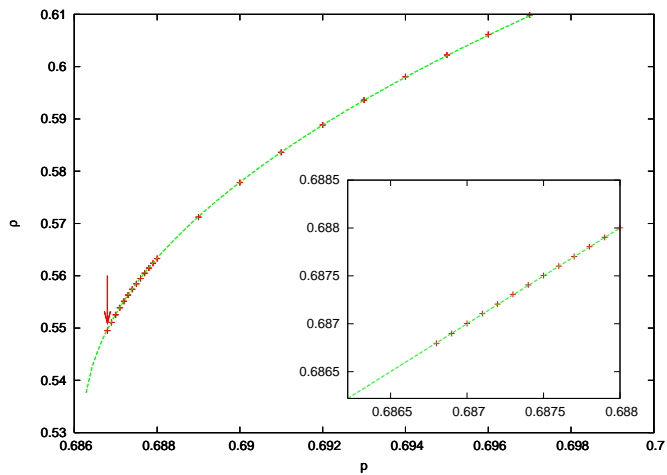,width=9cm} 
\caption{$k$-core density vs probability with $k=4$ for a four-dimensional sample of size $L=320$ and periodic boundary conditions. The arrow marks a discontinuous transition in the density that jumps to $\rho\approx .05$ at lower values of $p$. The data are fit in the region near the transition with the function $\rho(p) \simeq \rho(p_c)+b (p-p_c)^{1/2}$ but the actual transition probability (marked by the arrow) is above the $p_c$ estimated from the fit. Inset: rescaled plot of the data and the fit near the transition}
\label{dens_per}
\end{center}\end{figure}

We started the numerical investigations considering  hypercubic lattices with periodic boundary conditions (PBC).
In fig. (\ref{dens_per}) we plot the density of the $k$-core for a sample of size $L=320$, corresponding to $O(10^{10})$ sites.  
The density of the $k$-core has a discontinuous transition at a $p=.6869$, where it jumps from a high-density percolating phase to a low-density non-percolating phase $\rho \approx .05$. The behavior appears to be consistent with a singular behavior at the transition but  a careful study of the data in order to extract the critical $p$ and the exponent $\beta$ shows some inconsistencies. 
Indeed the curve seems to be fit at best with the exponent $\beta=1/2$ (the mean-field  Bethe-lattice value) but with a value of the critical probability $p\approx .6862$ definitively {\it lower} than that at which the transition is actually observed.
In order to assess if this behavior can be considered a finite-size effect we investigated the  $k$-core spatial correlation function.
\begin{figure}[htb]
\begin{center}
\epsfig{file=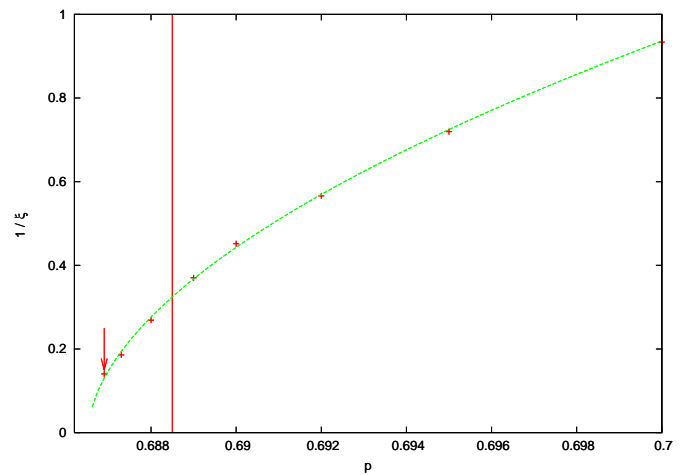,width=9cm} 
\caption{Inverse of the correlation length $\xi$ vs. probability for the same sample of fig. \ref{dens_per}. The arrow marks the point of  the actual transition. Although the behavior is consistent with a divergence at $p\approx .6862$, in all sample studied we never observed a correlation length at the actual transition exceeding $\xi\approx 8-10$, {\it i.e.} large but much smaller than the sample size $L=320$. The vertical line marks the value of the true critical probability $p_c$, see text.}
\label{corre}
\end{center}\end{figure}
In figure (\ref{corre}) we plot the inverse correlation length $\xi$ as a function of $p$ for the same sample, again  the plot is apparently consistent with a divergence of $\xi$ at a value of $p$ slightly lower than the value at which the transition is actually observed (marked with an arrow in fig. \ref{corre}).
In principle this could be a finite size effect but the problem is that the value of $\xi$ at the actual transition is large ($\xi \approx 8-10$) but {\it not comparable} with the size of the system ($L=320$).
Actually   we were not able to observe a correlation length bigger that ten lattice spacing {\it independently of the size of the system (up to $L=320$)} and it seems highly unlikely that this divergence drives the transition.
These features made us suspect that the the actual mechanism driving the transition is not the divergence of the correlation length but rather the nucleation of droplets of the low-density phase. 
Due to the periodic boundary conditions these nucleation centers are originally absent in the system but they appear as spatial fluctuations of the density when the correlation length is sufficiently big leading to the transition.
We tested this idea by putting  some nucleation center ({\it i.e.} empty hypercubes of size $l$) by hand in the sample  and checked that this procedure shifts the transition at higher values of $p$ although the size of the hypercubes ($l=20,40$) is small with respect to the size of the system and large with respect to the correlation length.
According to this interpretation the percolating phase is unstable and can be observed only because  nucleation centers in finite-size systems with periodic boundary conditions are extremely rare. 

In order to assess the validity of this interpretation and to determine whether there is a true percolation transition at higher values of $p$ we considered systems with  completely empty boundaries. 
These
boundary conditions guarantee that the system is completely isolated from the outside, as a consequence the
density at finite size is a {\it lower bound} to the density in the thermodynamic limit.
Furthermore it turned out that in this case numerical methods are rather safe for extrapolating the behavior of the
infinite system  at variance with the more delicate $k>d$ case where  finite-size effects are extremely large as mentioned above \cite{DeGregorio1}.
The possibility of choosing these boundary condition is a special feature of the case $k\leq d$ because otherwise
no site can resist culling if the boundaries of the hypercube are empty.

\begin{figure}[htb]
\begin{center}
\epsfig{file=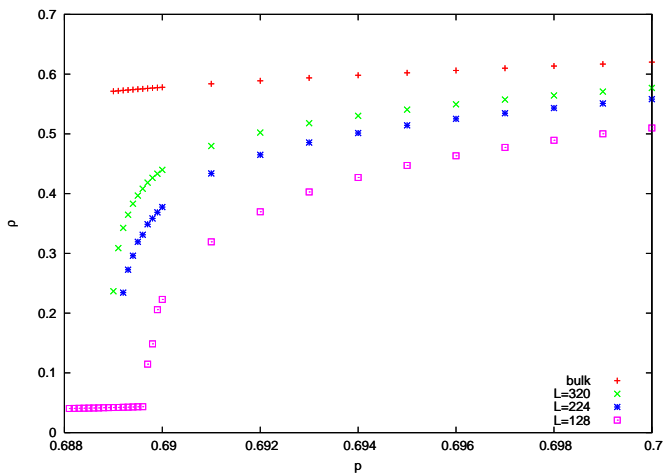,width=9cm} 
\caption{bulk density and total density vs. probability for different sample sizes, the average
total density at given size is a lower bound to the bulk density because of the empty boundary
conditions.}
\label{dens_tot}
\end{center}\end{figure}

In order to estimate the density at a given value of $p$ we generated lattices of increasing size.
In figure (\ref{dens_tot}) we plot the results for the {\it total density} $\rho_L$ at various sizes $L$. The total density $\rho_L$ is affected by the presence of the empty boundaries and {\it is an increasing
function of $L$}. The monotonicity property allows to safely conclude from fig. (\ref{dens_tot}) that there is indeed a discontinuous transition in the large $L$ limit.
We also measured the bulk density, 
{\it i.e.} the density of a smaller hypercube inside the sample whose boundaries are far enough from the surfaces. The bulk density provides a direct estimate of the density in thermodynamic limit and strengthens the conclusion that there is a discontinuous transition, see fig. \ref{dens_tot}. 
The transition probability decreases with the sample size and tends to a critical value $p_c=.6885(5)$ that was estimated through extrapolation.
We expect that the interface between the low-density phase on the boundaries and the high-density phase in the bulk penetrates more and more in the sample for $p\rightarrow p_c$, as a consequence in smaller systems the transition will occur at higher values of $p$. 
\begin{figure}[htb]
\begin{center}
\epsfig{file=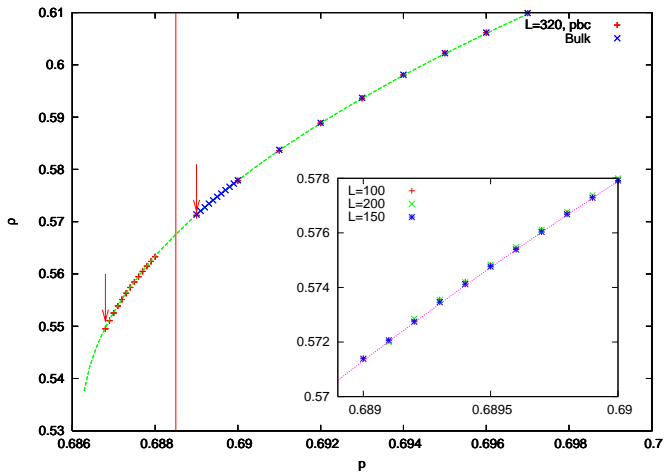,width=9cm} 
\caption{Density vs. probability for a sample with periodic boundary conditions and for the bulk of a sample with empty boundary condition with the same size $L=320$. The data and the fit for periodic boundary conditions are the same of in fig. \ref{dens_per}. The vertical line marks the estimated critical probability $p_c=.6885(5)$. The two arrows mark the position of the actual transition probabilities, the transition with PBC is in the unstable region, see text. Inset: magnified view of the bulk density near the transition computed on internal hypercubes of various sizes ($l=100,150,200$) inside a sample with $L=320$. }
\label{dens_bulk}
\end{center}\end{figure}
The percolation value $p_c=.6885(5)$ is larger than the actual value of the transition in the case of periodic boundary condition, {\it e.g.} $p=.6869$ for $L=320$,
in figure \ref{dens_bulk} we plot the bulk density and compare it with the density of the system in the case of periodic boundary conditions. For $p\geq p_c$ (marked by a vertical line in the plot) the two densities are equal  and we clearly see that the density in the case of periodic boundary conditions is the analytic continuation of the percolating-phase density in the unstable region.
We also verified that the correlation length of the bulk is the same  of the systems with periodic boundary conditions at the corresponding values of $p$, see fig. \ref{corre}.
By looking at figures \ref{dens_bulk} and \ref{corre} we immediately see that the density and correlation length are  regular at the estimated value of the real transition probability $p_c=.6885$. In particular using this value of $p_c$ we estimate  $\xi_c\approx 2.5$, while in the unstable phase we can observe $\xi$ up to ten lattice spacings.
We note that due to time and memory constraints, few lattices of the largest size $(L=320)$ can be studied, however the sample-to-sample fluctuations are small enough to ensure a good estimate the density, for instance  the bulk density estimates computed from hypercubes of different size $L=100,150,200$ do not show significant deviations, see inset of fig. \ref{corre}. 

\begin{figure}[htb]
\begin{center}
\epsfig{file=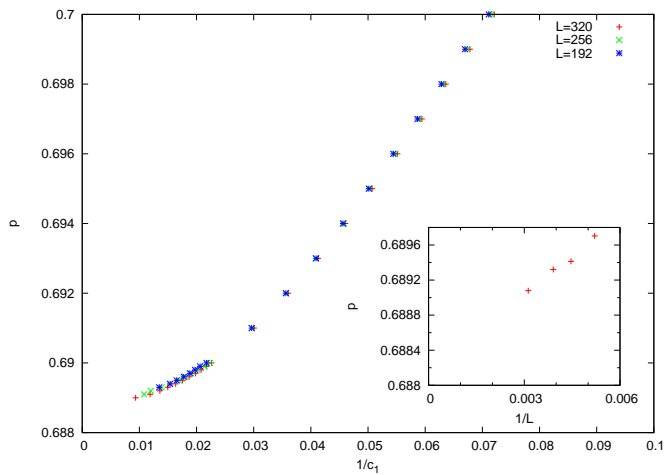,width=9cm} 
\caption{probability vs. inverse of the $1/L$ correction $c_1$ for different sample sizes, the data
are consistent with a divergence of the prefactors at the transition. Inset: probability at which
the total density is $\rho=.3<\rho_c$ as a function of the inverse of the size of the sample, it
tends to $p_c$ as the size of the system increases.}
\label{pref_pc}
\end{center}\end{figure}

The qualitative features of the finite size effect in the case of empty boundaries can be understood 
in the same way as in thermodynamical first-order transitions, {\it e.g.}  a ferromagnetic Ising model in a small positive field with the spins on the boundaries  forced to be negative.
As we noted above, the critical transition probability is shifted to higher values of $p$ and the true $p_c$ can be estimated through extrapolation (for instance for $L=320$ the transition is at $p=.689$, see \ref{corre}).
In general the finite size corrections to the density should scale as $1/L$, {\it i.e.}
\beq
\rho_L(p)=\rho(p)-{ 1\over L}c_1(p)+O({1 \over L^2}) 
\eeq
and figure \ref{dens_tot} suggests that the factor $c_1(p)$  diverges at $p_c$.
This factor is  determined by the density profiles $\rho(p,z)$ at distance $z$ from  an empty
surface:
\beq
c_1(p)=\int_0^{\infty} (\rho(p)-\rho(p,z))dz \ .
\eeq
Direct inspection of the profiles $\rho(p,z)$ shows that they are consistent with a divergence at the
transition, consistently with the expectation that the transition is determined by the 
 penetration deep inside the sample of the interface between the low-density and the percolating phases.
The precise nature of the divergence of $c_1(p)$  would require a more
detailed analysis which goes beyond the scope of this work.
In figure (\ref{pref_pc}) we plot the behavior of the inverse of $c_1$ and of the transition probability for
different sample sizes from which $p_c$ was estimated.

In conclusion the density of the $k$-core in four dimension with $k=4$ exhibits a discontinuous transition at $p_c=.6885(5)$
from a high-density percolating phase to a low-density non-percolating phase.
The transition loses its hybrid character with respect to the Bethe lattice case:
the density is regular at the transition and the correlation length is finite.
These results are most clearly seen numerically considering samples with empty boundary conditions, instead in finite size systems with periodic boundary conditions it is possible to follow the percolating phase in the unstable region $p<.6885(5)$. The behaviour of the unstable phase is consistent with a {\it pseudo}-transition at a lower probability $p\approx .686$. 
This pseudo-transition appears to have a hybrid character with pseudo-exponent $\beta=1/2$ and diverging correlation length but it cannot be observed because the unstable phase decays at $p\approx .6869$ through nucleation of low-density droplets determined by spatial fluctuations of the density. 

It would be very interesting to confirm these results through analytical methods.
Although the $1/d$ expansion \cite{Harris2} gives no hint about the observed disappearance of the mixed transition, maybe different approaches to perturbation theory around the Bethe solution \cite{MR,PS} could be able to see it, notwithstanding the possibility that non-perturbative effects should be taken into account.


\begin{thebibliography}{99}

\bibitem{Chalupa}  J. Chalupa, P. L. Leath, and G. R. Reich, J. Phys. C 12, L31 (1981).


\bibitem{Adler1}
J. Adler, Phys. A {\bf 171} 453 (1991)

\bibitem{Adler2}
J. Adler and U. Lev, Braz. Jour. of Phys. {\bf 33} 641 (2003)

\bibitem{Schon}
R. H. Schonmann, (1990) J. Stat. Phys. {\bf 58} 1239, 
 (1990)  Phys. A {\bf 167} 619, Ann. Probab. {\bf 20} 174 


\bibitem{Aizenman} M. Aizenman and J.L. Lebowitz, J. Phys. A, {\bf 21} 3801 (1988) 

\bibitem{Holroyd} A. Holroyd, Probab. Theory Relat. Fields 125, 195 (2003)


\bibitem{Cerf} R. Cerf and F. Manzo, Stochastic Processes Appl. 101, 69 (2002),
R. Cerf and N.M. Cirillo, Ann. Prob. {\bf 27} 1837 (1999) 


\bibitem{VanEnter}  A. C. D. van Enter, J. Stat. Phys. 48, 943 (1987).

\bibitem{DeGregorio1} P. De Gregorio, A. Lawlor, P. Bradley, and K. A. Dawson
Phys. Rev. Lett. 93, 025501 (2004)


\bibitem{Harris2} A. B. Harris and J. M. Schwarz
Phys. Rev. E 72, 046123 (2005)


\bibitem{Koguth} P. M. Koguth and P. L. Leath J. Phys. C. 14 529 (1981)
	

\bibitem{Schwarz} J. M. Schwartz, A. J. Liu and L. Q. Chayes, Europhys. Lett., 73 (4), pp.
560566 (2006)


\bibitem{jamm} C. S. O'Hern, S. A. Langer, A. J. Liu and S. R. Nagel, PHys. Rev. E {\bf 68}, 011306
(2003); L.E. Silbert, A.J. Liu and S. R. Nagel, cond-mat/0501616  

\bibitem{Harris1} A. B. Harris and T. C. Lubensky, J. PHys. A 16 L365 (1983), A. B. Harris, PHys.
Rev. B {\bf 28} 2614 (1983)





\bibitem{Medeiros} M. C. Medeiros and C. M. Chaves, Physica A {\bf 234}, 604 

\bibitem{Branco}
N. S. Branco and C. J. Silva, Int. J. Mod. PHys. C 10 (1999) 921

\bibitem{Kurtsiefer} D. Kurtsiefer, int. J. Mod. PHys. C 14, 529 (2003)


\bibitem{Cris} C. Toninelli, G. Biroli, and D. S. Fisher
Phys. Rev. Lett. 96, 035702 (2006), C. Toninelli and G. Biroli, cond-mat/0512335.

\bibitem{MR} A. Montanari and T. Rizzo, J. Stat. Mech. (2005) P10011. 

\bibitem{PS} G. Parisi and F. Slanina,  J.Stat.Mech. 0602 (2006) L003. 



\end{thebibliography}
\end{document}